\documentclass[runningheads]{llncs}
\usepackage{graphicx}
\usepackage{amsfonts}
\usepackage{amsmath}
\usepackage{makecell}

\usepackage{threeparttable}
\usepackage{multirow}
\usepackage{booktabs}
\usepackage{bbm}
\usepackage{bbold}
\usepackage{todonotes}
\usepackage{bbding}

\begin{document}
\title{Semi-supervised Pathology Segmentation \\ with Disentangled Representations
}

\author{Haochuan Jiang\inst{1}\Envelope
\and Agisilaos Chartsias \inst{1}
\and Xinheng Zhang\inst{2,3}
\and Giorgos Papanastasiou\inst{4}
\and Scott Semple\inst{5}
\and Mark Dweck\inst{5}
\and David Semple\inst{5}
\and Rohan Dharmakumar\inst{3,6}
\and Sotirios A. Tsaftaris\inst{1,7}}

\authorrunning{H. Jiang et al.}

\institute{School of Engineering, University of Edinburgh,  U.K., \email{haochuan.jiang@ed.ac.uk}
\and Department of Bioengineering, University of California, U.S.
\and Biomedical Imaging Research Institute, Cedars-Sinai Medical Center, U.S.
\and School of Computer Science and Electronic Engineering, University of Essex, U.K.
\and Center for Cardiovascular Science, University of Edinburgh, U.K.
\and Department of Medicine, University of California, U.S.
\and The Alan Turing Institute, U.K.}

\maketitle              

\begin{abstract}
Automated pathology segmentation remains a valuable diagnostic tool in clinical practice. However, collecting training data is challenging. 
Semi-supervised approaches by combining labelled and unlabelled data can offer a solution to data scarcity.  An approach to semi-supervised learning relies on reconstruction objectives (as self-supervision objectives) that learns in a joint fashion suitable representations for the task. Here, we propose Anatomy-Pathology Disentanglement Network (APD-Net), a pathology segmentation model that attempts to learn jointly for the first time: disentanglement of anatomy, modality, and pathology. The model is trained in a semi-supervised fashion with new reconstruction losses directly aiming to improve pathology segmentation with limited annotations. In addition, a joint optimization strategy is proposed to fully take advantage of the available annotations. We evaluate our methods with two private cardiac infarction segmentation datasets with LGE-MRI scans. APD-Net can perform pathology segmentation with few annotations, maintain performance with different amounts of supervision, and outperform related deep learning methods. 

\end{abstract}

\section{Introduction}\label{Sec:Introduction}
Deep learning models for automated segmentation of pathological regions from medical images can provide valuable assistance to clinicians. 
However, such models require a considerable amount of annotated data to train, which may not be easy to obtain. Pathology annotation (as opposed to anatomical), relies also on carefully detecting normal tissue areas for direct comparison.
It is therefore appealing to train pathology segmentors by combining the available annotated data with larger numbers of unlabeled images in a semi-supervised learning scheme.

A typical strategy to segment pathology is to first locate the affected anatomy, e.g. the myocardium for cardiac infarction~\cite{moccia2019development}, and use the detected anatomy to guide the pathology prediction. 
However, since anatomical annotations are not always available, recent methods cascade two networks: one segments the anatomy of interest, and the second one segments the pathology~\cite{morshid2019machine,pang2019modified,roth2019liver}. 

Although these methods achieve accurate segmentation, they are typically fully supervised and sensitive annotations numbers, as seen in our experimental results specified in Sec.~\ref{sec:exps}.
Semi-supervised learning is promising to solve the issue by engaging unlabelled images. 
Recently, disentangled representations, i.e. structured latent spaces that are shared between labeled and unlabeled data, have provided a solution to semi-supervised learning~\cite{chartsias2019disentangled,huang2018multimodal}.
These methods typically use specialized encoders to separate anatomy (a spatial tensor) and imaging information (a vector encoding image appearance) in medical image applications~\cite{chartsias2019disentangled}, while they involve unlabeled data through reconstruction losses. 

In this paper, inspired by~\cite{chartsias2019disentangled}, we propose the Anatomy-Pathology Disentanglement Network (APD-Net).
APD-Net constructs a space of anatomy (a spatial tensor), modality (a vector), and pathology  (a spatial tensor) latent factors. 
Pathology is obtained by an encoder, which segments the pathology conditioned on both the image and the predicted anatomy mask. 
We focus on segmenting myocardial infarct, a challenging task due to its size, irregular shape and random location. 
APD-Net is optimized with several objectives. Among others, we introduce a novel ratio-based triplet loss to encourage the reconstruction of the pathology region by taking advantage of the pathology factor. 
In addition, the use of reconstruction losses, that are made possible with disentangled representations~\cite{chartsias2019disentangled}, makes APD-Net suitable for semi-supervised learning.
Finally, we train with both predicted and real anatomy and pathology masks (in a \textit{Teacher-Forcing} strategy~\cite{williams1989learning}) to further improve performance.
Our major \textbf{contributions} are summarized as follows:
\begin{itemize}
	\item We propose a method for disentangled representations of anatomy, modality, and pathology;
	\item The disentanglement is encouraged with a novel ratio-based triplet loss;
	\item We also proposed the \textit{Teacher-Forcing} training regime combining different scenarios of real and predicted inputs to make full use of available annotations during optimization;
	\item APD-Net improves the Dice score of state-of-the-art benchmarks on two private datasets for cardiac infarction segmentation when limited supervision is present, whilst maintaining performance in the full annotation setting.  
\end{itemize}

\section{Related work}\label{Sec:RelatedWork}
\noindent\textbf{Pathology Segmentation:}
A classical approach to segment pathology is by cascading organ segmentation before segmenting the pathological region.
These two segmentors can be trained separately~\cite{morshid2019machine}, or jointly~\cite{pang2019modified,roth2019liver}.
In contrast to anatomy, pathology is small in sizes and irregular in shapes; thus, shape priors cannot be used.
To train models when masks are small, the Tversky loss~\cite{salehi2017tversky,roth2019liver} or similarly the focal loss in~\cite{abraham2019novel} have been proposed.
Our proposed model also cascades two segmentations, by using the initial anatomy prediction to guide the subsequent pathology segmentation.
However, we achieve this using disentangled representations to enable  semi-supervised learning. 

\noindent\textbf{Disentangled representations.}
The idea of disentangled representations is to decompose the latent space into domain-invariant spatial content latent factors (known as anatomy in medical imaging) and domain-related vector style ones (here referred to as modality)~\cite{huang2018multimodal,qin2019unsupervised}.  In medical image analysis, images are disentangled in anatomical and imaging factors for the purpose of semi-supervised segmentation \cite{chartsias2019disentangled}, image registration~\cite{qin2019unsupervised}, and classification~\cite{van2018learning}. Image reconstruction for semi-supervised segmentation was also investigated in~\cite{dey2018compnet} with a simpler disentanglement of the foreground (predicted anatomy masks) and the remaining background.
However, less effort has been placed on disentangling pathology. 
Some pioneering studies were conducted in~\cite{xia2020pseudo}, treating brain lesion segmentations as a pathology factor to synthesise pseudo-healthy images.
In this paper, we also adopt a segmentation of pathology as a latent factor and combine it with disentangled anatomy and modality~\cite{chartsias2019disentangled}, which enables the image reconstruction task for semi-supervised learning. 
While we are inspired by others~\cite{xia2020pseudo} who consider anatomy and pathology factors independently. This work is the first to learn them in a joint fashion.

\section{Methodology}\label{Sec:Methodology}
This section presents the APD-Net model. 
We first introduce relevant notations.
Then, we specify disentanglement properties (Sec.~\ref{sec:pathology_disentanglement}), detail the model architecture (Sec.~\ref{sec:architecture}), and finally present the learning objectives (Sec.~\ref{sec:losses}) with the joint training strategy depending on the different input scenarios (Sec.~\ref{sec:joint}).

\noindent\textbf{Notation:}
Let $X$, $Y_{ana}$, $Y_{pat}$ be sets of volume slices, and the associated anatomy and pathology masks, respectively.
Let $i$ be a sample. 
We assume a fully labeled pathology subset $\{x^i, y^i_{ana}, y^i_{pat}\}$, where $x^i \in X \subset \mathbb{R}^{H\times\ W \times 1}$, $y_{ana}^i \in Y_{ana} := \{0,1\}^{H \times W \times N}$, and $y^i_{pat} \in Y_{pat} := \{0, 1\}^{H \times W \times K}$. $N$ and $K$ denote the number of anatomy, and pathology masks respectively. $H$ and $W$ are the image height and width.
When $Y_{pat}=\emptyset$, it degrades to an unlabeled pathology set.\footnote{We only consider unlabeled pathology, assuming anatomy masks are available during training. Partial anatomy annotation is out of the scope of this paper.} 
Both anatomy and pathology sets are involved in a semi-supervised fashion to segment pathology.
This is achieved by learning a mapping function $f$ that estimates anatomy and pathology given an image $x^i$, i.e. $\{\hat{y}^i_{ana},\hat{y}^i_{pat}\} = f(x^i)$.

\subsection{Pathology Disentanglement}\label{sec:pathology_disentanglement}
\begin{figure}[t!]
    \centering
    \includegraphics[width=\linewidth]{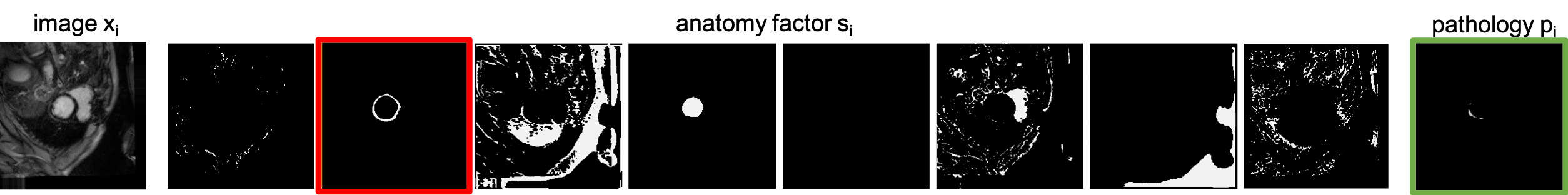}
    \caption{Visualising the disentanglement of the spatial (and binary) anatomy and pathology factors for a  LGE-MRI slice.}\label{fig:disentanglement}
\end{figure}

The main idea of APD-Net is to take an input image and decompose it into latent factors that relate to anatomy, pathology and image appearance (modality). This will allow inference of anatomical and pathology segmentations, whereas the disentanglement of the image appearance will enable image reconstruction that is critical to enable training with unlabeled images and semi-supervised learning. 
Herein we consider $C=8$ channels of anatomy factors and $n_z =8$ for modality factors as in~\cite{chartsias2019disentangled}, and $K=1$ (myocardial infarct).
$s^i$ and $p^i$ are obtained by softmax and sigmoid output activations respectively.
They are then binarised (per-channel) by $s^i -> [s^i+0.5]$ and $p^i -> [p^i+0.5]$, such that each pixel corresponds to exactly one channel. 
This binarisation encourages the produced anatomy factor to be modality-invariant. Finally, as in~\cite{chartsias2019disentangled} gradients are bypassed in the backward pass to enable back-propagation.

Fig.~\ref{fig:disentanglement} illustrates predicted anatomy and pathology factors for a cardiac infarct example. 
We make an intuitive distinction between the two factors in that the former only refers to healthy anatomical regions. Therefore, there is an overlap between the pathology factor and one or more anatomy channels. 
In this example (Fig.~\ref{fig:disentanglement}), the pathology (infarct in the green box) is spatially correlated with the myocardial channel (red box). 
Encoding pathology in the anatomy factor (i.e.\ entanglement of these two) is prevented, both through architecture design and with relevant losses. This will be detailed in the following sections.

\subsection{APD-Net Architecture}\label{sec:architecture}
\begin{figure}[t!]
	\centering
	\includegraphics[width=0.9\linewidth]{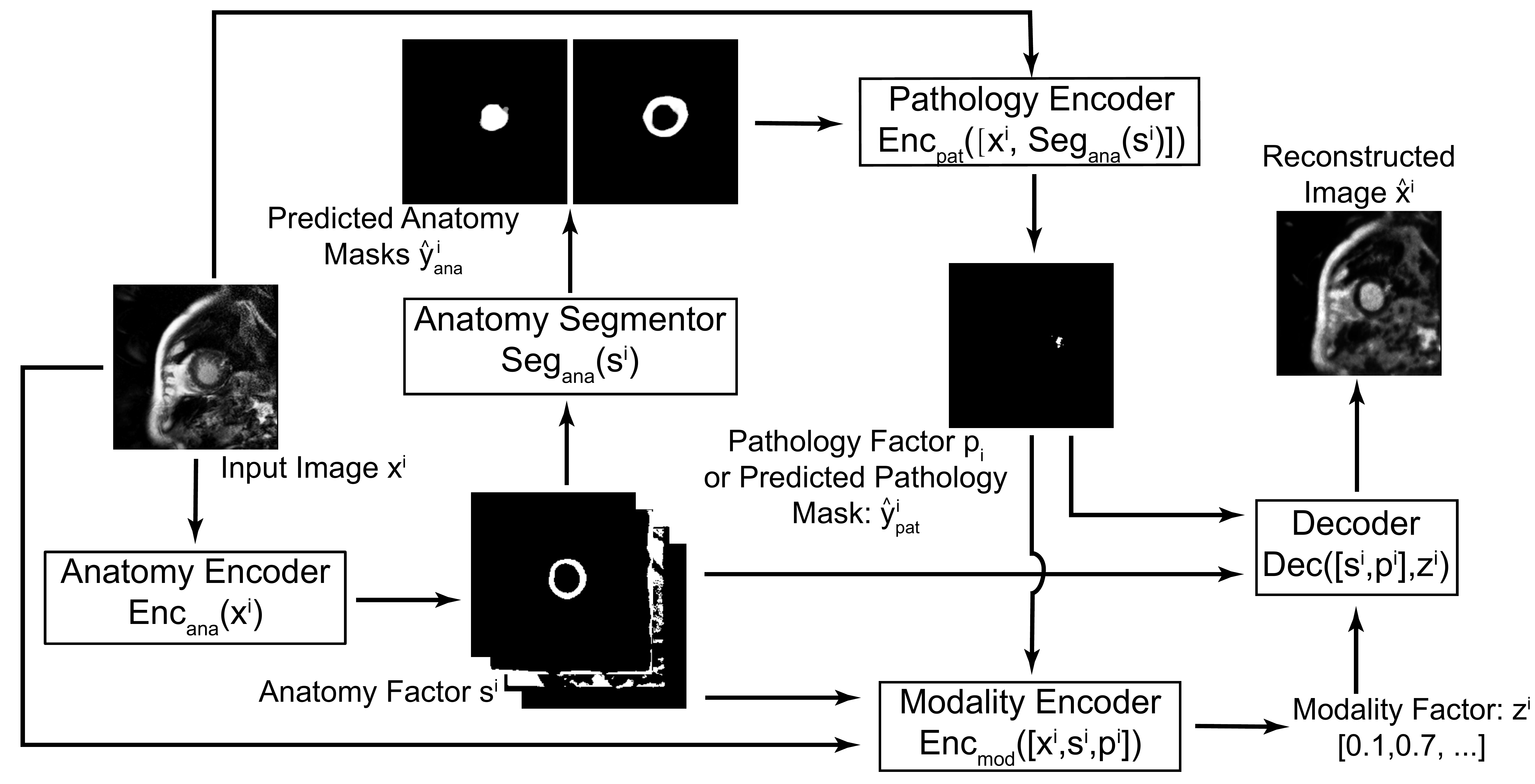}	
	\caption{Schematic of APD-Net. An image is encoded to anatomy factors using $Enc_{ana}$, and segmented with $Seg_{ana}$ to produce anatomical segmentation masks (in this case the myocardium and left ventricle). Combined with the input, the anatomy segmentation is used to segment the pathology with $Enc_{pat}$. Finally, given the anatomy, the pathology, and the modality factors from $Enc_{Mod}$, the decoder reconstructs the input. }\label{fig:APD-Net-Full}
\end{figure}

APD-Net, depicted in Fig.~\ref{fig:APD-Net-Full}, adopts modules from SD-Net~\cite{chartsias2019disentangled} including anatomy $Enc_{ana}$ and modality  $Enc_{mod}$ encoders, anatomy segmentor $Seg_{ana}$, and decoder $Dec$. 
They give $s^i$, $z^i$, the anatomy mask $\hat{y}^i_{ana}$, and the reconstructed image $\hat{x}^i$.

We introduce a pathology encoder $Enc_{pat}$ following the U-Net~\cite{ronneberger2015u} architecture. 
Given channel-wise concatenated $x^i$ and $\hat{y}^i_{ana}$, $Enc_{pat}$ produces $p^i$.\footnote{Note that $p^i$ is the same as $\hat{y}^i_{pat}$, i.e. the predicted pathology mask. We use $p^i$ for disentanglement and image reconstruction, and $\hat{y}^i_{pat}$ for pathology segmentation.}
Thus, APD-Net structurally resembles the cascaded segmentation scheme~\cite{pang2019modified}, enabling $Enc_{pat}$ to focus on specific regions to locate the pathological tissue.

Finally, the image decoder receives the concatenation of $s^i$, $p^i$, and $z^i$ to reconstruct image $\hat{x}^i$, enabling unsupervised training. With nulled pathology $p^i_0=\mathbb{0}$ (all elements are zero), a pseudo-healthy image ($\hat{x}^i_0$) is obtained~\cite{xia2020pseudo}.

\subsection{Individual Training Losses}\label{sec:losses}
APD-Net is jointly trained with losses including new supervised, unsupervised, and objectives selected from~\cite{chartsias2019disentangled,chartsias2019disentangle} for the task of pathology segmentation.

\noindent \textbf{Pathology Supervised Losses:}
Pathology manifests in various shapes. Thus, using a mask discriminator as a shape prior~\cite{chartsias2018factorised} is not advised. In addition, pathology covers a small portion of the image leading to class imbalance between the foreground and background. To address these shortcomings, inspired by ~\cite{roth2019liver,tran2019improving}, we combine Tversky\footnote{The Dice loss can be seen as a special case of the Tversky loss~\cite{salehi2017tversky} when $\beta=0.5$.}~\cite{salehi2017tversky} and focal loss~\cite{lin2017focal}. The Tversky loss is defined as follows: $\ell_{patT}=(\hat{y}^i_{pat} \odot y^i_{pat}) /[ \hat{y}^i_{pat} + y^i_{pat} + (1-\beta)\cdot(\hat{y}^i_{pat}-\hat{y}^i_{pat} \odot y^i_{pat})+\beta\cdot(y^i_{pat}-\hat{y}^i_{pat} \odot y^i_{pat})]$, where $\odot$ represents the element-wise multiplication.
The focal loss is defined as $\ell_{patF} = \sum_{H,W}[-y^i_{pat}(1-\hat{y}^i_{pat})^\gamma \log(\hat{y}^i_{pat})]$. 

\noindent \textbf{Pathology-Masked Image Reconstruction Loss:}
Typically image reconstruction is achieved by minimising the $\ell_1$ or $\ell_2$ loss between $x^i$ and $\hat{x}^i$.
However, due to the size imbalance between pathological and healthy regions, conventional full image reconstruction may ignore (average out) the pathology region. 
A simple, but effective solution, is to measure reconstruction performance on the pathology region by using the real pathology mask.
This is implemented by a masked reconstruction $\ell_1$ loss: $\ell_{M}= \frac{1}{H\times W \times C}\sum_{H,W,C}(\frac{\lambda_{pat}}{\lambda_{ana}}\cdot y^i_{pat} + \mathbb{1}) \cdot\|x^i-\hat{x}^i\|_1$.  $\mathbb{1}$ denotes a matrix with $y^i_{pat}$ dimensions, where all elements are ones.\footnote{The $\mathbb{1}$ matrix is added to ensure that no zero elements are multiplied with $\|x^i-\hat{x}^i\|_1$. Also, if $\lambda_{pat}=1$, the loss reduces to the $\ell_1$ loss.}

\noindent \textbf{Ratio-based Triplet Loss:}
However, this may not be adequate for accurate pathology reconstruction. We therefore penalise the model when pathology is possibly ignored by adopting a contrastive Triplet loss~\cite{schroff2015facenet}. 
This is defined as  $\max(m+d_{pos}-d_{neg}, 0)$, and minimizes the inter-class distance ($d_{pos}$) compared to the intra-class ($d_{neg}$) in deep feature space based on a margin ($m$).
We generate \emph{pseudo-healthy} images as negative examples, $\hat{x}^i_{0} = Dec(s^i, p^i_0, z^i)$, obtained by nulling the pathology factor, i.e.\ $p^i_0$.
The deep features are calculated as a new output attached to the penultimate layer of a reconstruction discriminator (inherent from the SD-Net~\cite{chartsias2019disentangle} for adversarial loss on $\hat{x}^i$) for the decoder output (denoted as $T$). 
By choosing $T(\hat{x}^i)$ as the anchor~\cite{schroff2015facenet}, positive and negative distances are calculated as $d_{pos} = \|T(\hat{x}^i)-T(x^i)\|_2^2$ and $d_{neg} = \|T(\hat{x}^i)-T(\hat{x}^i_0)\|_2^2$ respectively with corresponding samples $T(x^i)$ and $T(\hat{x}^i_0)$.

In practice, choosing a proper $m$ value is challenging~\cite{zakharov20173d}, particularly when the difference between the positive and the negative samples only lies in the small pathology region, e.g. in the cardiac infarction of Fig.~\ref{fig:disentanglement}. This will lead to an extremely small difference. Instead of optimizing the absolute margin $m$, we propose to alternatively minimize the Ratio-based triple loss (RT) defined as $\ell_{RT} = \max(r+\frac{d_{pos}}{d_{neg}}-1, 0)$.
The hyper-parameter $r$ represents the relative margin that the positive should be closer to the anchor than the negative.

\noindent \textbf{SD-Net Losses:}
We adopt optimization objectives from SD-Net~\cite{chartsias2019disentangled} and its multi-modal extension~\cite{chartsias2019disentangle} to the proposed framework. 
Specifically,  the anatomy supervision Dice loss,  the modality factor KL divergence, and the factor reconstruction loss are inherent from~\cite{chartsias2019disentangled}, while the adversarial loss on $\hat{x}^i$ is brought from~\cite{chartsias2019disentangle}. We refer SD-Net losses as $\ell_{SDNet}$.

\subsection{Joint Optimization with \textit{Teacher-Forcing} Training Strategy}\label{sec:joint}
A critical issue in cascaded architectures is that initial segmentation errors propagate and directly affect the second prediction. 
This should be taken into account particularly during training. We thus adopt the \textit{Teaching-Forcing} strategy~\cite{williams1989learning}, originally applied in RNNs by engaging real rather than predicted labels. 

In APD-Net, this strategy is applied on pathology segmentation $\hat{y}^i_{pat}$, modality factor estimation $z^i$, and reconstruction $\hat{x}_{i}$, which depend on real or predicted anatomy and pathology segmentations. Specifically, $\hat{y}^i_{pat}$ can be estimated using predicted anatomy masks,  $\hat{y}^i_{pat}=Enc_{pat}([x^i, Seg_{ana}(Enc_{ana}(x^i))])$, or real masks $\hat{y}^i_{pat}=Enc_{pat}([x^i, y^i_{ana}])$. Subsequently, the modality factor can be produced by the predicted pathology mask, $z^i=Enc_{mod}([x^i, Enc_{ana}(x^i), \hat{y}^i_{pat}])$, or the real pathology mask, $z^i=Enc_{mod}([x^i, Enc_{ana}(x^i), y^i_{pat}])$. Finally, real or predicted pathology contributes to reconstruction, $\hat{x}^i=Dec([Enc_{ana}(x^i), y^i_{pat}], z^i)$ and $\hat{x}^i=Dec([Enc_{ana}(x^i), \hat{y}^i_{pat}], z^i)$, respectively. 

We term the losses that involve the predicted, real anatomy mask, and real pathology mask as $\ell^{PA}$, $\ell^{RA}$, and $\ell^{RP}$ respectively.
We redefine each loss as a weighted sum $\ell=\lambda^{PA}\ell^{PA} + \lambda^{RA}\ell^{RA} + \lambda^{RP}\ell^{RP}$, where $\lambda^{PA}$, $\lambda^{RA}$, and $\lambda^{RP}$ are relative weights, and $\ell \in \{\ell_z, \ell_{RT},  \ell_{adv},  \ell_{M}, \ell_{KL}\}$.
Finally, the full objective is given by $\ell_{APD-Net} = \lambda_{patT}\ell_{patT} + \lambda_{patF}\ell_{patF}
		+ \lambda_{RT}\ell_{RT} + \lambda_{M}\ell_{M} + \ell_{SDNet}$. 

		

\section{Experiments}\label{sec:exps}
We evaluate APD-Net on pathology segmentation using the Dice score. 
Experimental setup, datasets, benchmarks, and training details will be detailed below.

\noindent \textbf{Data:}
We use two private cardiac LGE datasets acquired at the Biomedical Imaging Research Institute of the Cedars-Sinai Medical Center (\emph{Data1}) and the Center for Cardiovascular Science of the University of Edinburgh (\emph{Data2}),
which have been approved by data ethics committees of the respective providers.
Both datasets contain annotations of the myocardium and myocardial infarct. \emph{Data1} involves 45 subjects (36 used for training) and $224 \times 224$ dimension. 
\emph{Data2} consists of 26 (mixed healthy and pathology) subjects (20 used for training), and $192\times 192$ dimension. 

\noindent \textbf{Benchmarks:}
We compare APD-Net with three benchmarks on infarct segmentation. 
\textit{U-Net (masked):} A U-Net trained on images $x^i$ masked by the ground truth myocardium mask $y^i_{ana}$~\cite{moccia2019development}. Masking here facilitates training, reducing the task to finding only infarcted myocardial pixels.
\textit{U-Net (unmasked):} The U-Net is trained on images $x_i$ without masking. This is more challenging since now the U-Net implicitly has to find infarct pixels from the whole image.
\textit{Cascaded U-Net}~\cite{pang2019modified}: this trains one U-Net to segment the myocardium (with 100\% supervision) and another to segment the infarct after masking the input image with the predicted myocardium (varying the number of available annotations).\footnote{\textit{U-Net (masked / unmasked)} and \textit{Cascaded U-Net} are optimized with full supervision using Tversky and focal losses, and penalized as defined in the \textbf{Training details}. In reality, \textit{U-Net (masked)} is not a good choice since manual myocardial annotations are not always available at inference time.}

\noindent \textbf{Training details:}
Training penalties are set to: $\lambda_{patT}$=1, $\lambda_{patF}$=1.5, $\lambda_{RT}$=1, $\lambda_{M}$=3.
Weights for different optimization scenarios are: $\lambda^{PA}$=1, $\lambda^{RA}$=0.7, and $\lambda^{RP}$=0.5.
The relative margin for $\ell_{RT}$ is $r$=0.3.
Other hyper-parameters include $\beta$=0.7 in $\ell_{patT}$, $\gamma$=2 in $\ell_{patF}$, and the SD-Net weights defined in~\cite{chartsias2019disentangled}.
Due to the small data size, we do not specify validation sets. 
All models are trained for fixed 100 epochs, and results reported below contain averaged Dice scores and standard deviation on test data of two different splits.\footnote{Code will be available at \url{https://github.com/falconjhc/APD-Net} shortly.}

\begin{table}[t!]
\centering
\caption{Performance evaluation for \emph{Data1} and \emph{Data2}. We report test Dice scores (with standard deviation in subscript calculated by summarizing all the involved volumes) on infarct segmentation with varying infarct supervision (\% infarct). }\label{tab:performance}
\begin{tabular}{c|c|c|c|c}
\hline
Dataset - \% infarct & {\textit{U-Net (unmasked)}} & {\textit{Cascaded U-Net}} & APD-Net & {\textit{U-Net (masked)}} \\ \hline
\emph{Data1}-13\%                                                                  
& $5.4_{8.1}$                                           
& $36.2_{17.9}$              
& ${45.3_{14.4}}$             
& $57.3_{13.5}$   \\ 

\emph{Data1}-25\%                                                                  
& $6.5_{8.4}$                                           
& $39.4_{15.6}$              
& ${46.4_{12.8}}$             
& $57.3_{13.2}$   \\ 

\emph{Data1}-50\%                                                                  
& $26.2_{13.8}$                                           
& $37.1_{13.8}$              
& ${46.7_{11.5}}$             
& $66.4_{9.0}$   \\ 

\emph{Data1}-100\%                                                                 
& $34.6_{15.3}$                                           
& $36.4_{17.9}$              
& ${47.4_{17.0}}$             
& $65.6_{10.3}$   \\ \hline

\emph{Data2}-13\%                                                                
& $11.5_{24.4}$                                           
& $25.6_{23.9}$              
& ${40.0_{27.0}}$             
& $21.5_{25.6}$   \\ 

\emph{Data2}-25\%                                                                  
& $36.9_{29.7}$                                           
& $35.8_{23.4}$              
& ${40.5_{26.7}}$             
& $46.7_{25.6}$   \\ 

\emph{Data2}-50\%                                                                  
& $15.0_{14.5}$                                           
& $34.4_{24.2}$              
& ${38.4_{17.0}}$             
& $49.8_{26.4}$   \\ 

\emph{Data2}-100\%                                                               
& $16.5_{16.2}$                                           
& $33.4_{16.4}$              
& ${38.9_{15.9}}$             
& $45.7_{28.1}$   \\ \hline
\end{tabular}
\end{table}
\subsection{Results and Discussion}
\noindent\textbf{Semi-supervised Pathology Segmentation:}
We evaluate the APD-Net performance in a semi-supervised experiment by altering the pathology supervision percentage, as seen in Table~\ref{tab:performance} respectively for the two datasets.
For clarity we omit anatomy segmentation results, which are approximately 78\% for both Cascaded U-Net and the proposed APD-Net for \emph{Data1} and 64\% for \emph{Data2}.

Infarct segmentation is a challenging task, and thus all results of \emph{Data1} and \emph{Data2} present relatively high standard deviation, in agreement with previous literature~\cite{karim2016evaluation}.
In \emph{Data1}, APD-Net consistently improves the Dice score of infarct prediction for all amounts of supervision, compared with both the \textit{Cascaded U-Net} and the \textit{U-Net (unmasked)}. 
Furthermore, the performance of APD-Net on small amounts of pathology labels is equivalent to the fully supervised setting.

\begin{figure}[t!]
	\centering
	\includegraphics[width=0.95\textwidth]{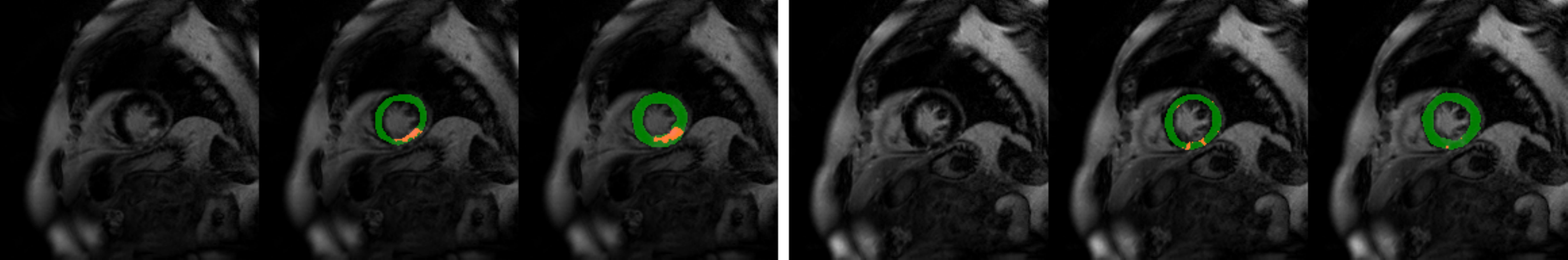}	
	\caption{Segmentation examples from \emph{Data2}. The two panels show a good and failure infarct segmentation case. 
	For each sample, the left image shows the input, and the next two overlay real and predicted myocardium and infarct respectively.
} 
	\label{fig:data2_examples}
\end{figure}

Segmenting pathology in \emph{Data2} is harder, as evidenced by the lower mean and higher standard deviation obtained from all methods. This could be due to the supervised methods overfitting to the smaller dataset size. APD-Net, however, overcomes this issue with semi-supervised training, and outperforms the \textit{U-Net (masked)} at 13\% annotations. While at 100\% annotations, APD-Net achieves the equivalent Dice as 13 \% but reduces the standard deviation.
More importantly, APD-Net outperforms the \textit{Cascaded U-Net} in all setups demonstrating the benefit of image reconstruction (see also ablation studies later).
Examples of correct and unsuccessful segmentations from APD-Net can be seen in Fig.~\ref{fig:data2_examples}, where the existence of sparsely-distributed annotations (right panel of Fig.~\ref{fig:data2_examples}) negatively affects supervised training.

\noindent\textbf{Ablation Studies:}
We evaluate the effects of critical components including the pathology-masked image reconstruction, disentanglement, teacher-forcing, and the ratio-based triplet loss with 13\% and 100\% infarct annotations on \emph{Data1}.
To evaluate disentanglement, we remove the modality encoder, and allow the anatomy factor to encode continuous values, $s^i \in [0,1]^{H\times W\times C}$. 
The results presented in Table~\ref{tab:ablation} show that canceling any of the ablated components hurts segmentation (except for the masked reconstruction, which at 100\% performs as the proposed APD-Net). 
In particular, reducing annotations at 13\% further decreases performance of the ablated models.

\begin{table}[t!]
\centering
\caption{Ablation studies on \emph{Data1} on two infarct annotation levels (\% infarct): no mask reconstruction ($\lambda_M=0$); 
no disentanglement (w.o. Disent.); no teacher-forcing strategy ($\lambda^{RA}=\lambda^{RP}=0$); and no ratio-based triplet loss  ($\lambda_{RT}=0$). }\label{tab:ablation}
\begin{tabular}{c|c|c|c|c|c}
\hline
\% annotations 
& $\lambda_M=0$ 
& w.o. Disent.
& $\lambda^{RA}=\lambda^{RP}=0$ 
& $\lambda_{RT}=0$ & APD-Net (proposed) \\ \hline
13\%
&$41.4_{15.2}$
&$14.9_{8.3}$
&$40.7_{12.4}$
&$38.8_{14.9}$
&$45.3_{14.4}$\\ \hline
100\%
&$47.7_{15.9}$
&$18.4_{16.6}$
&$44.5_{12.9}$
&$40.3_{10.7}$
&$47.4_{17.0}$\\ \hline
\end{tabular}
\end{table}

\begin{figure}[t!]
	\centering
	\includegraphics[width=0.95\textwidth]{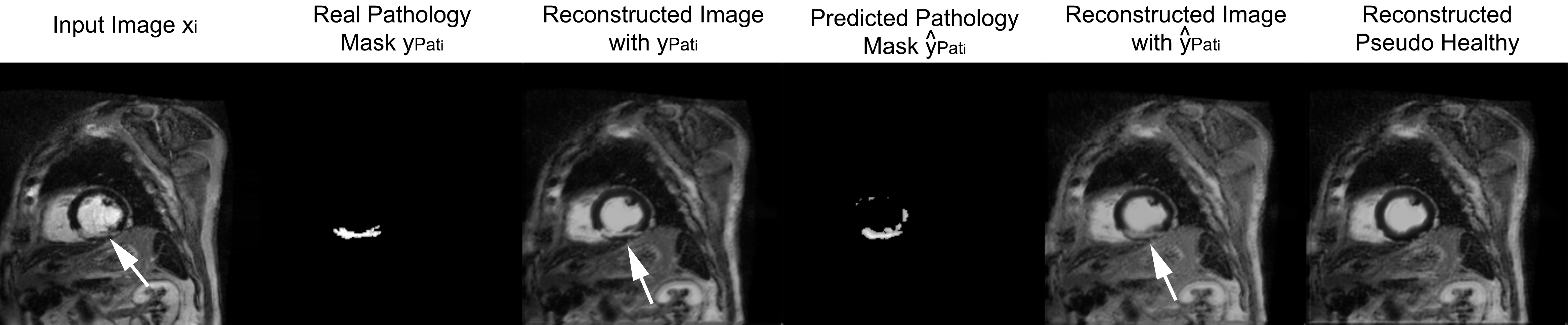}	
	\caption{Reconstruction visualizations with real,  predicted pathology masks, and pseudo-healthy. White arrows point at infarct regions. }\label{fig:visualization}
\end{figure}
Fig.~\ref{fig:visualization} depicts the effects the disentangled pathology factor on the reconstructed image. Arrows indicate the infarct region, evident when using either the real or predicted pathology to reconstruct the image. In contrast, the infarct is missing when the pathology factor is nulled $p_0^i$, producing a pseudo-healthy image. Qualitatively, the synthetic image of the proposed APD-Net is similar to the one presented in~\cite{chartsias2019disentangled}.
The difference between the reconstructed and pseudo-healthy images is driven by the pathology factor. It is enhanced by the ratio based triplet loss that is essential for the desired pathology disentanglement.

\section{Conclusions and Future Work}
In this paper, we proposed the Anatomy Pathology Disentanglement Network (APD-Net) disentangling the latent space into anatomy, modality, and pathology factors. 
Trained in a semi-supervised scenario with reconstruction losses enabled by disentangled representation learning and joint optimization losses, APD-Net is capable of segmenting the pathology region effectively when partial pathology annotations are available. 
APD-Net has shown promising results in pathology segmentation using partial annotations and improved performance compared to other related baselines in the literature.

However, APD-Net still follows the cascaded pathology segmentation strategy that can propagate errors from the first to the second segmentation, which is not solved fundamentally even with the \textit{Teacher-Forcing}.
Diseases that deform the anatomical structure (e.g., brain tumour) cannot be predicted by cascading.
In addition, we only tested the proposed APD-Net in myocardial infarct, where the pathology manifests as high-intensity regions within the myocardium.
As future work, we aim to explore direct pathology segmentation methods without predicting the relevant anatomy. 
Meanwhile, we plan to investigate extensions of the current APD-Net that are more general and do not restrict to myocardial infarct, while also engaging multi-modal images that would offer complementary anatomical information.
Finally, we will test our method on public datasets with more examples to further validate pathology segmentation performance.

~\\
\noindent\textbf{Acknowledgement:} This work was supported by US National Institutes of Health (1R01HL136578-01). This work used resources provided by the Edinburgh Compute and Data Facility (http://www.ecdf.ed.ac.uk/).
S.A. Tsaftaris acknowledges the Royal Academy of Engineering and the Research Chairs and Senior Research Fellowships scheme. 

\bibliographystyle{splncs04}
\bibliography{miccai}

\end{document}